\documentclass{svjour3}                     
\smartqed  
\usepackage{graphicx}
\usepackage{psfrag}
\usepackage{epsfig}
\usepackage{amsmath}
\usepackage{color}
\usepackage{soul}
\usepackage[normalem]{ulem}
\journalname{Experiments in Fluids}
\begin{document}

\title{The effect of dead time on randomly sampled power spectral estimates
}

\author{Preben Buchhave     \and
        Clara M. Velte         \and
        William K. George
}


\institute{Preben Buchhave \at
            Intarsia Optics\\
            S{\o}nderskovvej 3\\
            3460 Birker{\o}d, Denmark\\
            \email{preben.buchhave@get2net.dk}
           \and
           Clara M. Velte \at
            Department of Mechanical Engineering\\
            Technical University of Denmark\\
              Nils Koppels All\'{e} Bldg. 403 \\
              2800 Kgs. Lyngby, Denmark\\
              \and
              William K. George \at
           Department of Mechanical and Aerospace Engineering\\
           Princeton University\\
           Princeton, NJ 08544\\
            and\\
            Department of Aeronautics\\
            Imperial College London\\
            South Kensington Campus\\
            London SW7 2AZ\\
}

\date{Received: date / Accepted: date}

\maketitle

\begin{abstract}
We investigate power spectra of a randomly sampled stationary stochastic signal, e.g. a spatial component of a turbulent velocity. We extend the methods of previous authors that basically assumed point or delta function sampling by including features characteristic of real measurement systems. We consider both the effect on the measured spectrum of a finite sampling time, i.e., a finite time during which the signal is acquired, and a finite dead time, that is a time in which the signal processor is busy evaluating a data point and therefore unable to measure a subsequent data point arriving within the dead time delay.
\end{abstract}

\section{Introduction}
\label{intro}

Estimation of power spectra from randomly sampled data is still a matter of concern, and several strategies have been proposed to obtain the best (most accurate and least noisy) and fastest (shortest data processing) algorithms. The foundation was laid already in the 1950ies by e.g. Blackmann and Tukey~\cite{1} and by Shapiro and Silvermann~\cite{2} and further developed by e.g. Gaster and Roberts in the 1970ies and 1980ies~\cite{3,4b,5}. The work by Gaster and Roberts in particular has served as a reference for much of the subsequent development of power spectral estimation of especially laser Doppler anemometer (LDA) measurements. However, due to the limitations in computing power at that time, these investigations basically assumed point sampling in space and time, and many issues connected with the processing of signals from a system with a finite measurement volume (MV) were unresolved.

In the present work we shall present the results of an investigation of the effects of a finite measurement volume resulting in a finite sampling time, during which the detector and signal processor are busy processing the current data point, and in a finite dead time, in which the system is demobilized and unable to receive a new data point, on the measured power spectra from an instrument collecting randomly arriving data, for example a laser Doppler anemometer. Initially, we focus on processes in which the sampling and sampled processes can be assumed statistically independent. However, throughout the work with computer generated LDA data we apply the so-called residence time corrected spectral estimators that compensate for the fact that the sample rate is correlated with the magnitude of the velocity measured by a burst-mode LDA.

In this paper we review the ideal case of random sampling of a stationary random process represented by a string of delta functions arriving at random times. In the limit of infinite data sets we retrieve the mean square constant offset known from previous investigations, e.g. Gaster and Roberts~\cite{3,4b,5}. We first review the complications arising due to a finite record length and from digital processing. Then, to mimic the spatial/temporal filtering introduced by all practical measurement instruments, we extend the theory to include a signal that is measured as an average over a top hat sampling function and show that the effect on the power spectrum is a filtering effect with a sinc-squared transfer function. Finally, we consider the effect of a finite dead time, during which the instrument locks out new acquisitions, and show that the resulting spectrum is severely biased when the mean sampling interval approaches the dead time of the detector and signal processor. We illustrate the problems with analytical expressions and plot the effect of filtering and dead time on a typical von K\'{a}rm\'{a}n-like power spectrum as it occurs in turbulent fluid flow. We also compare the analytical expressions with results of data processing of a set of data generated by random sampling of a random process with a von K\'{a}rm\'{a}n power spectrum and show excellent agreement with the analysis.

\section{The system under investigation} \label{sec:2}

\begin{figure*}[!h]
  \center{\includegraphics[width=0.5\textwidth]{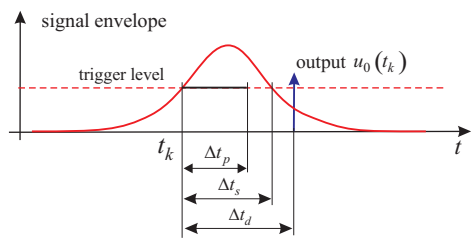}}
\caption{The sampling process.} \label{fig:Fig1}
\end{figure*}

We shall consider a measurement system consisting of a detector and signal processor whose function it is to obtain data from a random process, say, $u(t) = \overline{u} + u'(t)$, where $\overline{u}$ is the temporal mean and $u'(t)$ is the fluctuating part. In the following we shall refer to the carrier of the information, e.g. a Doppler modulated electronic pulse, as the ``signal'' and to the result of the measurement, $u(t)$, as the ``velocity''. We consider a burst type processor, where a signal is detected by a burst detector when the signal level exceeds a certain threshold. The burst ends after a time $\Delta t_s$, which we shall denote the residence time or transit time. During the burst, the signal is digitized and processed to provide one numerical velocity output for each burst. We assume the processing of a signal burst starts immediately after the burst detection and continues through a processing time $\Delta t_p$, which ideally would be equal to $\Delta t_s$, but could also be a fixed value, smaller than $\Delta t_s$, determined by the function of the signal processor. Finally, we may consider a case where the processor needs some time to transfer the data point to the data processor (i.e. computer) before it is ready for the next measurement. We denote the total time from the threshold burst detection until the processor is again ready for a measurement, the dead time, $\Delta t_d$. The total output from the signal processor for each burst is the measured velocity, $u_0(t_k)$, the residence time, $\Delta t_s$, and the time of arrival, $t_k$. See Figure~\ref{fig:Fig1} for an illustration of the detection model.

We assume a sampling function, $g(t)$, which assigns the measurement to the particular arrival times $t_k$, and describes the detector response and the statistical properties of the sampling process. The measured signal, $u_0(t)$, can then be written as a continuous function of time (see, e.g., George \textit{et al.}~\cite{11})
\begin{equation}
u_0(t) = u(t)g(t).
\end{equation}

\subsection{Ideal sampling}

Ideally, a measurement of a time dependent process should occur at a single point in space and time. We may represent random sampling of such a process by a sampling function, which is a string of delta-functions placed at random times, say $t_k$:
\begin{equation}
g(t) = \frac{1}{\nu} \delta(t-t_k),\qquad t_k\,\mathrm{random}
\end{equation}
Note that we have normalized by $\nu$, the average number of samples per unit time in order to make the sampling function $g(t)$ dimensionless, see~\cite{4}.

\begin{figure*}[!h]
  \center{\includegraphics[width=0.5\textwidth]{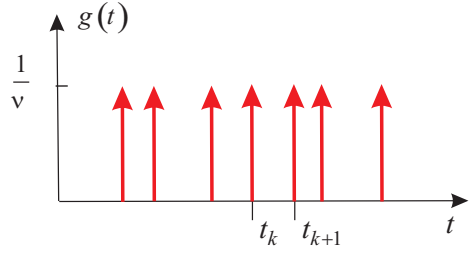}}
\caption{Ideal random sampling represented by a string of delta functions weighted by $1/\nu$ at random times $t_k$.} \label{fig:Fig2}
\end{figure*}

Figure~\ref{fig:Fig2} is a sketch of a typical $g(t)$.

Thus the measured signal becomes
\begin{equation}\label{eq:3}
u_0(t) = u(t) \cdot \frac{1}{\nu}\delta(t-t_k),\qquad t_k\,\mathrm{random}
\end{equation}

\subsubsection{First order statistics}
Since our sampling process is a stationary random process, we can use the equivalence of time and ensemble averaging to write:
\begin{eqnarray}
\overline{g(t)} & = & \lim_{T \rightarrow \infty} \frac{1}{T}\int_{0}^{T} g(t)\,dt \nonumber \\
& = & \lim_{T \rightarrow \infty} \frac{1}{T}\int_{0}^{T} \sum_{k=1}^N \frac{1}{\nu} \delta(t-t_k)\,dt \nonumber \\
& = &  \lim_{T \rightarrow \infty} \frac{1}{\nu T} \sum_{k=1}^N \int_{0}^{T} \delta(t-t_k)\,dt
\end{eqnarray}
where $N$ is the number of samples in time $T$ and random.  But the integral over each delta function is unity  and the expected value of $N$, say $\langle N \rangle$, is just $\nu T$;
so
\begin{equation}
\overline{g(t)}  = 1
\end{equation}
Thus the normalization chosen for $g(t)$ gives it a mean value of unity.

Assuming statistical independence between the measurement process and the sampled process, the mean value of the sampled velocity is then
\begin{equation}
\overline{u_0(t)}= \overline{g(t)} \cdot \overline{u(t)} = \overline{u}   \label{eq:meanuo}
\end{equation}

\subsubsection{Second order statistics: the autocovariance and power spectrum}

\noindent {\bf The autocovariance function}

\noindent The sampling function autocovariance $\langle g(t) g(t+\tau) \rangle$ can be expressed as (see Appendix C in Velte~\cite{4}, George \textit{et al.}~\cite{11}):
\begin{equation}
\overline{g(t) g(t+\tau)}= 1 + \frac{1}{\nu}   \delta(\tau) \label{eq:gcorrelation}
\end{equation}

If the measurement process and the sampling process are statistically independent, the correlation for the measured signal can be written as:
\begin{eqnarray}
 \overline{u_0(t) u_0 (t+\tau)} &  = & \overline{u(t)g(t)\cdot u(t+\tau)g(t+\tau)} \nonumber \\
&  = & \overline{u(t)u(t+\tau)} \cdot \left [1 + \frac{1}{\nu} \delta(\tau) \right ] 
\end{eqnarray}

\noindent {\bf The Power Spectrum}

\noindent The power spectrum is found by using the Wiener-Khinchine theorem:
\begin{eqnarray}
S_{u_0}(f) &=& FT \left \{\overline{u_0(t) u_0 (t+\tau)} \right \} \nonumber \\
&=& FT \left \{ \overline{u(t) u (t+\tau)} \cdot  \overline{g(t) g (t+\tau)}\right \} \nonumber \\
&=& FT \left \{ \overline{u(t) u (t+\tau)}\right \} \otimes FT \left \{ \overline{g(t) g (t+\tau)}\right \} \nonumber \\
&=& S_u(f) \otimes S_g(f)
\end{eqnarray}
where FT denotes the Fourier transform. $S_u(f)$ corresponds to the true spectrum of $u$, and $S_g(f)$ is the Fourier transform of the sampling function ACF:
\begin{equation}
S_g(f) = \delta(f) + \frac{1}{\nu}. \label{eq:Sg}
\end{equation}
Thus the spectrum of $u_0(t)$ is given by:
\begin{equation}
S_{u_0}(f)= S_{u}(f) + \frac{\overline{u^2}}{\nu}. \label{eq:idealspectrum}
\end{equation}
Or if one splits the instantaneous velocity into mean and fluctuating parts:
\begin{equation}
S_{u_0}(f)= S_{u'}(f) + \overline{u}^2\delta (f) + \frac{\overline{(\overline{u}+u')^2}}{\nu}.
\end{equation}

\begin{figure*}[!h]
  \center{\includegraphics[width=0.5\textwidth]{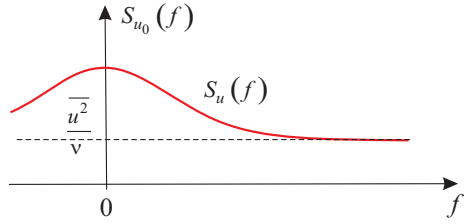}}
\caption{The measured spectrum consists of the true spectrum and constant spectral offset.} \label{fig:Fig4}
\end{figure*}

The measured spectrum is identical to the true spectrum plus a constant offset that does not include any spectral information, see Figure~\ref{fig:Fig4}. This term is a consequence of the way we calculate the power spectrum.  The contribution of the mean velocity is normally eliminated in processing, so only the contribution of the variance remains.  It has been suggested  by Gaster and Roberts~\cite{3} and others (us included~\cite{11,4,6b,12}) that this constant offset can be subtracted, or eliminated by processing (e.g.~\cite{4,12}).  Unfortunately as demonstrated in the following sections, the problem is somewhat more complicated because of real signal processing concerns.  The problem of finite time of measurement is well-known, but often the consequences are not realized. Further, the problems of dead-times and measuring time do not seem to have been addressed at all. These problems form the main point of this paper.

\subsection{Finite Measurement Records}

\subsubsection{Spectrum via Correlation Function}

Since practical measurement records are by necessity bound in time, the analysis is always performed using finite windows. The resulting spectrum is thus a convolution of the spectrum of the original signal with the Fourier transform of a finite time window function. Exactly which window depends on the method of processing.  The most common (especially in LDA applications) has been to compute the ACF first using the so-called `time-slot approximation', then transform it to obtain the spectrum.  If the correlation function is simply truncated at the largest time lag possible, say $(-T,T)$, the resulting transformation is:
\begin{eqnarray}
S_T(f) & = & \int_{-T}^{T} e^{-i 2 \pi f \tau}~C_u(\tau)\, d\tau \nonumber \\
& = & \int_{-\infty}^\infty S(f-f_1) W_{tophat}(f_1)\,df_1  \label{eq:finitetimecorrelationmethod}
\end{eqnarray}
where
\begin{equation}
W_{tophat}(f) = \int_{-T}^{T} e^{-i 2 \pi f \tau} df = T \left[ \frac{\sin(2\pi f T)}{2\pi f T} \right] = T \,\mathrm{sinc} (2 \pi f T)
\end{equation}
Note that in the limit as $T \rightarrow \infty$, $W_{tophat}(f) \rightarrow \delta(f)$, so the infinite domain spectrum is recovered. The spectrum of our randomly sampled signal of infinite record length in equation~(\ref{eq:idealspectrum}) can be substituted into equation~(\ref{eq:finitetimecorrelationmethod}) to obtain the finite time estimator as:
\begin{equation}
S_{u_0,\,T}(f) = \int_{-\infty}^\infty S_u(f-f_1) W_{tophat}(f_1)\,df_1  ~ + ~ \frac{\overline{u^2}}{\nu} \label{eq:finitetimespectrum}
\end{equation}

\subsubsection{The direct spectral estimator}
An alternative to the time-slot approximation to spectral analysis is to use the direct spectral estimator proposed by George \textit{et al.} (1978). For an \textit{infinite time-interval}, the so-called direct method is mathematically equivalent to the correlation method which goes by way of the ACF and the Wiener-Khinchine theorem. However, the direct method is the preferred method of the two mainly for two reasons; The slotting method re-introduces aliasing by arranging the data into equi-spaced slots and the direct method is often computationally more efficient in array based math software. We next display the direct method applied to delta function sampling.

To see this, consider first the Fourier transform of $u_0(t)$ {\it in the sense of generalized functions} (c.f.~\cite{Lumley1970}) of the sampled signal, which is {\it defined} to be the following:
\begin{equation}
\tilde{u}_0(f) = FT_{gf}\{u_0(t)\} \equiv \lim_{A \rightarrow \infty} \int_{-\infty}^\infty e^{-i 2 \pi f t} u_0(t) g_A(t) \,dt
\end{equation}
where $g_A(t)$ makes the integral converge and $\lim_{A \rightarrow \infty} g_A(t) = 1$. It is straightforward to show that if $u(t)$ is a stationary random process, then the Fourier coefficients are uncorrelated at different frequencies and
\begin{equation}
\langle \tilde{u}_0^*(f) \tilde{u}_0(f') \rangle \,df \,df' = S_{u_0}(f) \delta(f'-f) \,df \,df'
\end{equation}
where $S_{u_0}(f)$ is exactly the spectrum given in equation~(\ref{eq:idealspectrum}).

But we never have an infinite record length over which to compute the transform, so George \textit{et al.}~\cite{11} (by analogy with continuous signal analysis) suggested using the {\it finite time transform} given by:
\begin{equation}
\tilde{u}_{0,T}(f) =  \int_{-T/2}^{T/2} e^{-i 2 \pi f t}   u_0(t) \, dt
\end{equation}
(Note that we have shifted the origin for the beginning of the record to be $-T/2$ to avoid a phase shift, but since we will multiply the finite Fourier transform by its complex conjugate, this was not really necessary and can be abandoned with the practical algorithm.)
The digital implementation of this has been discussed in~\cite{10}, but for now note that the finite record spectral estimator can be computed from this as:
\begin{equation}
\hat{S}_{u_{0},T}(f) = \frac{\langle \tilde{u}_0^*(f) \tilde{u}_0(f) \rangle}{T}
\end{equation}
where the averaging is necessary since each  transform product is random (like the signal itself).  From this point on the process is exactly analogous to standard signal processing with equi-spaced samples (except that the FFT cannot be used for randomly sampled data).  And in fact the effect of the finite record length on $\hat{S}_{u_0}(f)$ is exactly the same and can be shown quite readily to be given by:

\begin{equation}
\hat{S}_{u_{0,T}}(f) = \int_{-T}^{T} e^{-i 2 \pi f \tau} \overline{u_0(t)u_0(t+\tau)} \left [1 - \frac{|\tau|}{T} \right ]\, d\tau
\end{equation}
Thus this direct estimator has yielded a spectrum which is also convolved with a window function; namely,
\begin{equation}
\hat{S}_{u_{0,T}}(f)  = \int_{-\infty}^\infty S_{u_0}(f_1) W_{tr}(f-f_1)\, df_1
\end{equation}
where this time the window function is given by:

\begin{equation}
W_{tr}(f) = \int_{-T}^T e^{-i 2 \pi f \tau} \left [1 - \frac{|\tau|}{T} \right ]\, d\tau = T \left[\frac{\sin (\pi f T)}{\pi f T} \right]^2
\end{equation}
This is the familiar Bartlett or triangle window, which has two advantages over the top-hat above:  first it does not produce negative side-lobes, and second it falls off as $f^{-2}$ instead of $f^{-1}$, thereby producing less spectral leakage. (Note that in spite of the latter advantage it is still necessary in turbulence to use additional window functions to get the very highest frequencies, since turbulence spectra in the dissipation range roll off much faster that $f^{-2}$.) However, the most important advantage of the direct estimator is that by avoiding the discretization in time-lag of the time-slot approximation, it really has produced an un-aliased spectrum for randomly arriving samples.

\section{Real signals}

Real measurements require some finite time, the measurement time or processing time, $\Delta t_p$, to be executed, see Figure~\ref{fig:Fig1}. The effect on the measured data is that the real quantity to be measured is somehow filtered by the measurement process. The maximum processing time is the transit time or residence time, $\Delta t_s$. In addition, the signal processor may require some time to transfer a data point and recover before it is ready for the next measurement, the dead time, $\Delta t_d$. As will be shown in the following, the effect of processing time and dead time can significantly modify the ideal random sampling statistics by averaging the underlying signal over the processing time and by eliminating lags smaller than the dead time.

\subsection{The effects of processing time and signal averaging effects}
\begin{figure*}[!h]
  \center{\includegraphics[width=0.5\textwidth]{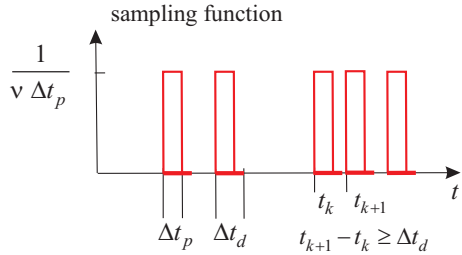}}
\caption{Top hat sampling function with dead time.} \label{fig:Fig6}
\end{figure*}

Similarly to the ideal sampling using delta functions, we can describe the sampling process by a top hat window function of width $\Delta t_p$ and height $1/(\nu \Delta t_p)$, see Figure~\ref{fig:Fig6}. Note that this will approach the case of ideal sampling with delta functions in the limit as $\Delta t_p \rightarrow 0$.

The top hat sampling function may be represented by
\begin{equation}
H_{\Delta t_p}(t) = \frac{1}{\Delta t_p} [h(t + \Delta t_p/2) - h(t - \Delta t_p/2)]
\end{equation}
where the factor $1/\Delta t_p$ calibrates time averages of the measured signal and $h(t)$ is the Heaviside function.  This defines a rectangular top hat function of unity integral and width $\Delta t_p$. The top hat function may be initiated at the arrival time $t=t_k$, a condition determined by the $g$-function we used before. However, the exact location of the interval about the arrival time is not important; a shift just results in a phase factor, which disappears in the power spectrum.

The detector and signal processor are further assumed to cause a simple averaging of the true signal during the processing time, $\Delta t_p$. (Note that the averaging time may depend on the measurement system and processor and could be as large as $\Delta t_s$, but generally for LDA is smaller and therefore here simply referred to as the processing time $\Delta t_p$.) Thus the signal to be measured, say $u_{\Delta t_p}(t)$ becomes:
\begin{eqnarray}
u_{\Delta t_p}(t) & =  & \frac{1}{\Delta t_p} \int_{t- \Delta t_p/2}^{t + \Delta t_p/2} u(t')  dt' \\
& = &  [ u(t) \otimes H_{\Delta t_p}(t)]  \nonumber
\end{eqnarray}
since averaging over the interval $\Delta t_p$ corresponds to convolving the true signal with the sampling function. Using our g-function from before, the measured signal at time $t_k$ becomes:
\begin{eqnarray}
u_{0,\Delta t_p}(t) & =  & g(t) ~u_{\Delta t_p}(t) \label{eq:upmeas} \nonumber \\
& = &  g(t)~ [ u(t) \otimes H_{\Delta t_p}(t)]  \nonumber \\
& = & g(t) \cdot \frac{1}{\Delta t_p} \int_{t- \Delta t_p/2}^{t + \Delta t_p/2} u(t')  dt'
\end{eqnarray}
This can be interpreted as a filtered signal sampled at random times

\subsubsection{The spectrum of $u_{\Delta t_p}$}
The correlation function can be computed as before, but in this case, because of the convolution in computing $u_{\Delta t_p}(t)$, it is easier to compute the spectrum first.  The Fourier transform of $u_{\Delta t_p}(t)$ {\it in the sense of generalized functions}, say $\tilde{u}_{\Delta t_p}(f)$,  is readily computed to be:
\begin{equation}
\tilde{u}_{\Delta t_p}(f) = \tilde{u}(f) ~  \mathrm{sinc} (\pi f \Delta t_p),
\end{equation}
since the Fourier transform of a convolution is the product of their Fourier transforms.  It follows immediately that the filtered and ideally sampled spectra are related by:
\begin{equation}
S_{u_{\Delta t_p}}(f) = S_u(f) ~ \mathrm{sinc}^2(\pi f \Delta t_p)  \label{eq:SDeltatp}
\end{equation}

\subsubsection{The spectrum and correlation of the sampled signal}
From equation~(\ref{eq:upmeas}) it follows immediately that $\overline{u_{0,\Delta t_p}(t) u_{0,\Delta t_p}(t + \tau)}$ is given by:
\begin{equation}
C_{u_{0,\Delta t_p}}(\tau) = \overline{u_{0,\Delta t_p}(t) u_{0,\Delta t_p}(t + \tau)} = \overline{g(t)g(t+\tau)}\cdot \overline{u_{\Delta t_p}(t)u_{\Delta t_p}(t+\tau)}
\end{equation}
But we know both correlation functions from equations~(\ref{eq:gcorrelation}) and (\ref{eq:idealspectrum}).

The spectrum, say $S_{u_{0,\Delta t_p}}(f)$, is then:
\begin{equation}
S_{u_{0,\Delta t_p}}(f) = S_g(f) \otimes S_{u_{\Delta t_p}}(f),
\end{equation}
which we also know from equations~(\ref{eq:Sg}) and (\ref{eq:SDeltatp}). Substitution yields immediately:
\begin{equation}
S_{u_{0,\Delta t_p}}(f) = \frac{\overline{u^2_{0,\Delta t_p}}}{\nu} +S_u(f) \cdot \mathrm{sinc}^2 (\pi f \Delta t_p).
\end{equation}
where
\begin{equation}
\overline{u^2_{0,\Delta t_p}} = \int_{-\infty}^\infty S_u(f-f_1) \cdot \mathrm{sinc}^2(\pi f_1 \Delta t_p) \,df_1
\end{equation}
The sinc-squared factor is a transmission function due to the filtering of the true signal taking place during the acquisition of each data point over the measurement time, $\Delta t_p$.

\subsection{Dead time effects}

\begin{figure*}[!h]
  \center{\includegraphics[width=0.5\textwidth]{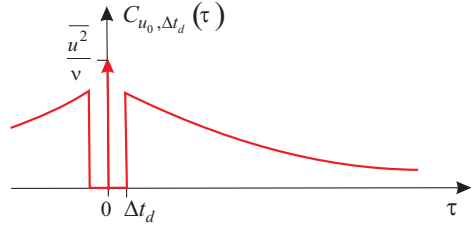}}
\caption{Measured ACF with dead time effect.} \label{fig:Fig7}
\end{figure*}

To see the effect of the dead time and the missing lags, $|\tau|<\Delta t_d$, we can revert to correlation space and introduce the lost lags  by removing them from the covariance function as shown in Figure~\ref{fig:Fig7}.

Alternatively the zero value could be replaced by some other value (say the smallest lag measured, or if known independently, the mean square value).  Note that at first glance this might appear to be the same as deleting the zero value for the time slot approximation considered earlier, but there is an important difference.  Here the smallest non-zero lag is determined by the hardware, not by the choice of time slots, and is thus unavoidable.

We introduce dead time by eliminating lags $|\tau|<\Delta t_d$:
\begin{equation}
C_{u_{\Delta t_p, \Delta t_d}}(\tau) = C_{u_{\Delta t_p}}(\tau) \cdot [1 - h(\tau + \Delta t_d) + h(\tau -\Delta t_d) ]
\end{equation}
The measured power spectrum is then:

\begin{equation}\label{eq:40}
S_{u_0,\Delta t_p, \Delta t_d} (f) = \frac{\overline{u^2_{0,\Delta t_p, \Delta t_d}}}{\nu_0} + \left [ S_{u}(f) \cdot \mathrm{sinc}^2 (\pi f \Delta t_p) \right ] \otimes \left [\delta (f) - 2 \Delta t_d ~\mathrm{sinc} (2 \pi f \Delta t_d) \right ]
\end{equation}
where $\nu_0$ is the reduced sample rate due to dead time\footnote{Reduction of sample rate: Let the original sample rate with no dead time be $\nu$. Then the number of samples in time $t$ is $n = \nu t$. The probability of $n$ samples in time $t$ with mean number of samples $\overline{n}$ (Poisson):

$$P(n) = \frac{e^{-n} \overline{n}^n}{n!}$$

or with $\overline{n} = \nu t$: $P(n) = \frac{e^{-\nu t} (\nu t)^n}{n!}$. The probability that no event occurs in time $\Delta t_d$ is then:

$$P(0) = e^{-\nu \Delta t_d}$$

But $P(0)$ is also the probability that the next sample will occur after $\Delta t_d$. Thus the rate of samples occurring after $\Delta t_d$, the reduced sample rate $\nu_0$, is $\nu_0 = \nu e^{- \nu \Delta t_d}$.}, $\nu_0 = \nu e^{-\nu \Delta t_d}$.
Point sampling, $\Delta t_p = 0$, but with dead time, is described by
\begin{equation}
S_{u_0, \Delta t_d} (f) = \frac{\overline{u^2_{0,\Delta t_d}}}{\nu_0} + S_{u}(f) \otimes \left [\delta (f) - 2 \Delta t_d ~\mathrm{sinc} (2 \pi f \Delta t_d) \right ]
\end{equation}

We now present some results of theoretical analysis and computer simulations. These calculations apply to what we consider a typical measurement situation. We take as our turbulence model a von K\'{a}rm\'{a}n spectrum with a low break point at 100\,Hz and a high break point at 3\,kHz. The mean sample rate for the measurements is varied between 1\,kHz and 20\,kHz. The measurement volume is adjusted to give dead times between 0 and 50\,$\mu$s. We average the computer generated spectra over typically 100 records, each with a record length of 1\,s. The spectral window is thus adequate to give sufficient resolution and negligible spectral leakage.

\begin{figure*}[!h]
\begin{minipage}{0.5\linewidth}
  \center{\includegraphics[width=1.0\textwidth]{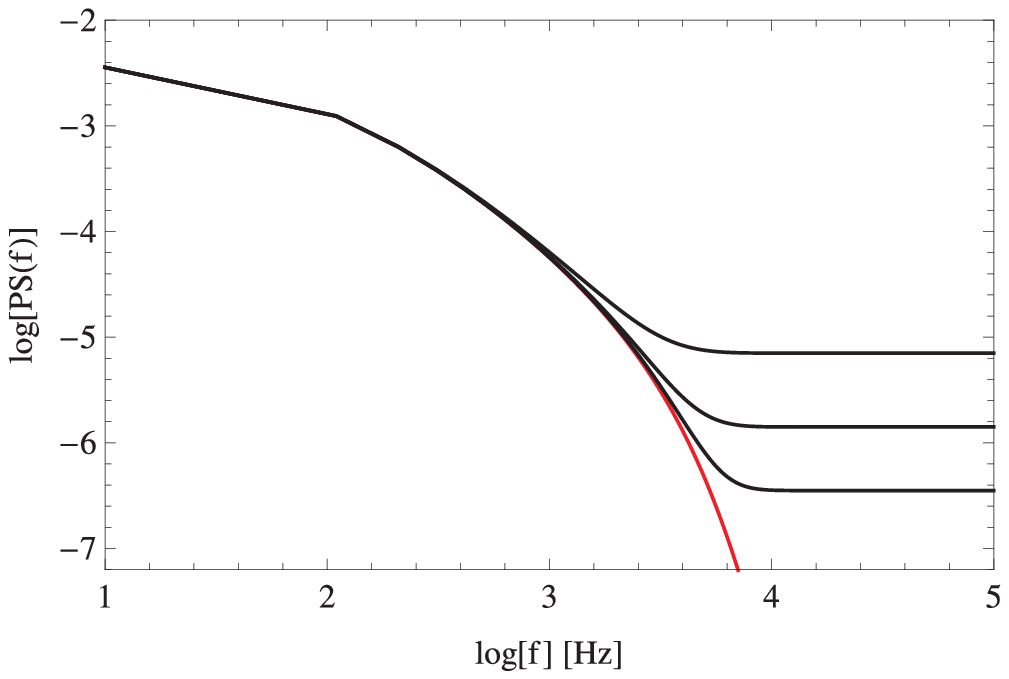}}
\end{minipage}\hspace{0.5cm}
\begin{minipage}{0.5\linewidth}
  \center{\includegraphics[width=1.0\textwidth]{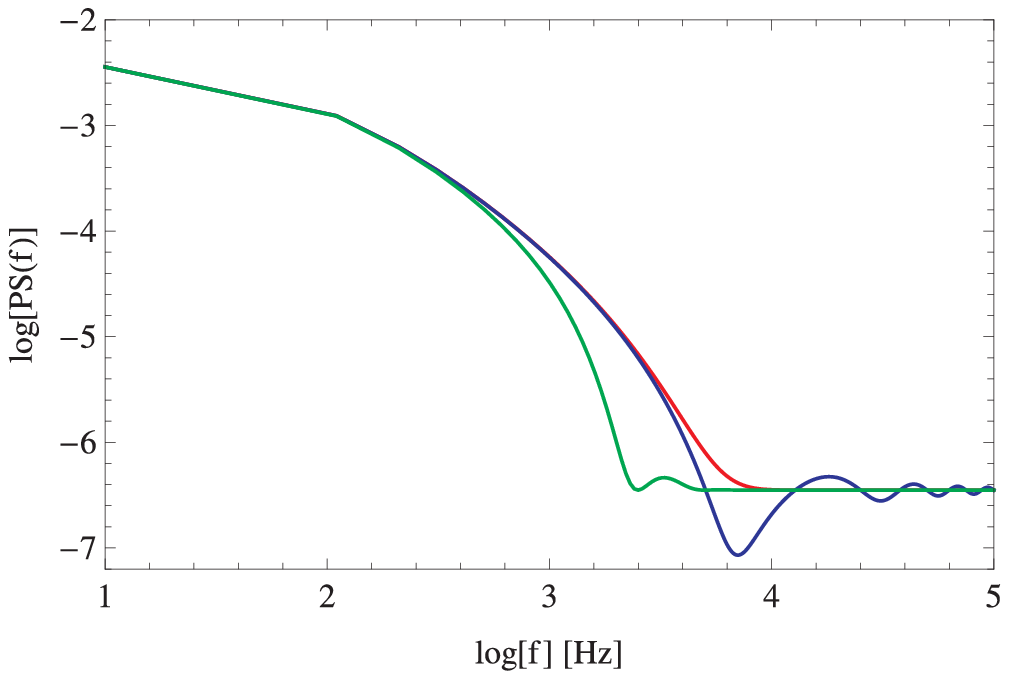}}
\end{minipage}
\caption{The effect of dead time on a von K\'{a}rm\'{a}n power spectrum. \textbf{Left:} Red: von K\'{a}rm\'{a}n model spectrum with infinite data rate, black: effect of varying data rate (1000\,Hz, 5000\,Hz, 20000\,Hz). \textbf{Right:} Red: von K\'{a}rm\'{a}n model with finite average sampling rate, blue: effect of dead time ($\Delta t_d = 0.00005$\,s), green: effect of filtering ($\Delta t_p = 0.0005$\,s).} \label{fig:Fig8}
\end{figure*}

In Figure~\ref{fig:Fig8} we illustrate the effect of varying the sampling rate $\nu$ with $\Delta t_d$ set to zero (left figure) and the effects of filtering and finite dead time in the case of finite average sampling rate (right figure) on a von K\'{a}rm\'{a}n model power spectrum.

\begin{figure*}[!h]
  \center{\includegraphics[width=1.0\textwidth]{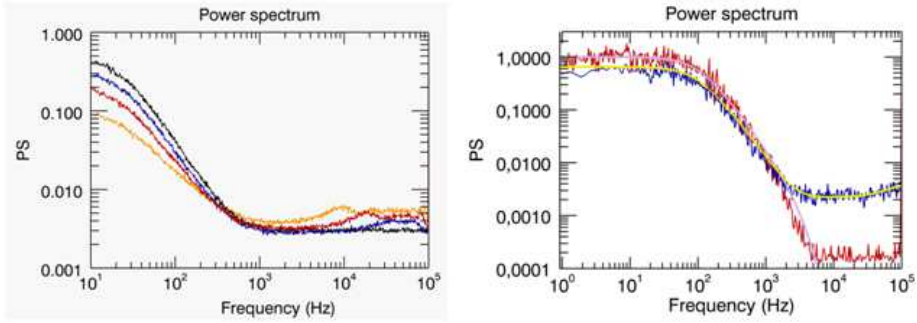}}
\caption{Effect of dead time on the processing of a simulated von K\'{a}rm\'{a}n spectrum. \textbf{Left:} Black: $\Delta t_d = 0\,s$. Blue: $\Delta t_d = 2\cdot10^{-6}\,s$. Red: $\Delta t_d = 4\cdot10^{-6}\,s$. Orange: $\Delta t_d = 8\cdot10^{-6}\,s$. \textbf{Right:} Light blue: von K\'{a}rm\'{a}n filter. Yellow: Analytical solution with dead time. Blue: Computer generated data with dead time, randomly sampled with $\nu =40.6\,kHz$. Red: Computer generated data, regular sampling at $100\,kHz$.} \label{fig:Fig9}
\end{figure*}

In Figure~\ref{fig:Fig9} we display the power spectrum of a set of data generated in a computer by random sampling of a velocity signal with a von K\'{a}rm\'{a}n spectrum as in Figure~\ref{fig:Fig8}. An equidistantly sampled primary velocity signal, $u_{primary}$, is initially generated using the FFT from a series of evenly distributed random frequency values, which is filtered in frequency with the von K\'{a}rm\'{a}n spectrum. The random arrival times are extracted from the primary velocity time series using a Poisson sampling process, $t_a = \mathrm{Poisson} \left \{ u_{primary}(t),\mu \right \}$ where $\mu$ is an adjustable parameter ensuring that the Poisson process provides primarily zeros or ones. The randomly sampled velocity data are subsequently extracted from the primary velocity signal, $u(t_a) = u_{primary}(t_a)$. To mimic the laser Doppler anemometer, the corresponding residence times are computed as $t_r(t_a) = d_{MV}/|u_{primary}(t_a)|$ where $d_{MV}$ is the diameter of the measurement volume. The temporal resolution is ultimately limited by the resolution of the primary time series.

The sampling process, being modulated by the instantaneous velocity magnitude, also includes simulation of a fixed detector dead time. The spectra are evaluated with the so-called residence time corrected power spectral estimators~\cite{11,10} by the direct method. Note that in Figure~\ref{fig:Fig9}, the spectra are significantly affected by dead time: The spectral power is reduced at low frequencies, while the spectral offset is increased a high frequencies. Also evident is the oscillation introduced by the sinc-term in the dead time response at the highest displayed frequencies. The left hand side shows the effect of varying the dead time, $\Delta t_d$. The right hand side shows the primary von K\'{a}rm\'{a}n model spectrum and the analytical model~(\ref{eq:40}) with an added constant offset. Also shown are the computer generated data, randomly sampled and regularly sampled.

\subsubsection{Dead time effects on white noise}

We can also study the effect of dead time on pure Poisson sampling, $\Delta t_s = 0$, of a constant signal with spectrum $\frac{a^2}{\nu_0}\delta (f)$, where $a^2$ is a constant mean square value. The spectrum then reduces to
\begin{equation}
S_{0} (f) = \frac{a^2}{\nu_0}\left [ 1 - 2 \Delta t_d ~\mathrm{sinc} (2 \pi f \Delta t_d) \right ]
\end{equation}
in agreement with previous studies of the effect of dead time on photon counting, e.g., Zhang \textit{et al.}~\cite{6}.

\section{Conclusion}

The problems with real randomly sampled signals are many, but only the most fundamental ones were considered here: namely that no instrument can measure instantaneously, but must average over some finite time to produce a realization. The measurement time (or processing time), $\Delta t_p$, is the time required for the measurement to be executed, i.e., the time over which the measurement is filtered. The dead time, $\Delta t_d$, is here defined as the measurement time plus the time that the processor may require for data transfer and to recover before the next measurement can be acquired.

The effects of the processing time and the dead time have been analyzed on a typical velocity power spectrum, closely described by a von K\'{a}rm\'{a}n spectrum. The effect of the signal processor on the measured velocity values was described as an averaging over the processor time, and it was shown that the result is a sinc-squared frequency transfer function cutting off the high frequency end of the spectrum.

The dead time, during which the processor cannot register new measurements, was shown to result in the (filtered) power spectrum being convolved with a sinc-function whose width depends on the extent of the dead time. This addition will typically introduce a clearly visible `dip' and oscillation in the spectrum at frequencies around and above the probe volume cut-off. The analysis was compared to the power spectra obtained from simple, but realistic, computer generated data, displaying excellent agreement with the analysis. As the dead time is further increased, the spectra display even more significant distortions such as a redistribution of power across frequency that are observed to severely bias the spectrum even at the lowest frequencies.

The special case of the laser Doppler anemometer is more problematical, since the dead time varies with each acquired data point and the sampled process is generally dependent on the process being sampled. Also, different commercially available processors display somewhat different behavior and their exact functioning is generally not disclosed to the user. Further, particle interference in the measuring volume will affect the residence time distribution, which however can be provided in the measuring process by correctly working processors. A detailed model for dead time in LDA is therefore beyond the scope of the current work but will be treated in a separate paper.

\end{document}